\documentstyle[12pt,aasms4]{article}

\begin{document}
\received{ }
\accepted{  }
\journalid{ }{ }

\slugcomment{   }

\lefthead{Barrado y Navascue\'es et al. }
\righthead{Stellar activity in coeval clusters}


\title{Stellar Activity in Coeval Open Clusters: Praesepe and
The Hyades}

\author{David Barrado y Navascu\'es\altaffilmark{1,2,3},
 John R. Stauffer\altaffilmark{3}} 
\affil{Harvard--Smithsonian Center for Astrophysics,
       60 Garden St., Cambridge, MA 02138, USA\\
dbarrado@cfa.harvard.edu, jstauffer@cfa.harvard.edu}

\and

\author{Sofia Randich}
\affil{Osservatorio Astrofisico di Arcetri, Largo
E. Fermi 5, I-50125, Firenze, Italy\\
randich@arcetri.astro.it}  

\altaffiltext{1}{MEC/Fulbright fellow.}
\altaffiltext{2}{Fellow at the Real Colegio Complutense, 26
Trowbridge St.,  Cambridge, MA 02138, USA.}
\altaffiltext{3}{Visiting Astronomer, Multiple Mirror Telescope.
Mt. Hopkins Observatory.}

\begin{abstract}

Randich and Schmitt [1995, A\&A 298, 115] found that the coronal activity
of solar-type and low mass stars in Praesepe is significantly lower than that
of stars in the Hyades cluster.  This result is quite surprising since the
Hyades and Praesepe have approximately the same age and metallicity
and are often thought to have originated in the same Giant Molecular 
Cloud complex.  We have carried out several tests in order to
find a possible explanation for this result.  We have measured
radial velocities  of two groups of Praesepe stars (a dF-dK
sample and a dM sample) and have measured H$\alpha$ as a chromospheric
activity index for the dM sample.  Based on analyses of these
data, we conclude that the Praesepe catalog used in the
X-ray analysis does not contain a significant number of non-members,
and thus that membership problems do not seem to be the cause of
the  Randich and Schmitt result.  The comparison of the
H$\alpha$ equivalent widths for the M dwarfs in Praesepe with those
in the Hyades indicates that, at least for stars in
this mass range, the Praesepe stars are as active or more active
than their Hyades counterparts.  The similarity of 
chromospheric emission allows us to reject differences in the
rotational velocity distribution as the origin of the
dissimilar Lx luminosity functions.

We have also analyzed a few ROSAT PSPC pointings of Praesepe
in order to obtain a new and independent estimate of
the X-ray luminosities and upper limits
for a small sample of Praesepe members. This analysis suggests 
that the previous ROSAT/PSPC analysis produced slightly optimistic X-ray upper
limits; however, the differences between the
  old and new upper limits are not large enough 
to explain the dichotomy in the X-ray properties of Praesepe
and the Hyades.  Therefore, our
examination of the available data does not provide  a 
clear reason to explain why the X-ray luminosity functions of the two
clusters are different.
Part of the explanation could be found in the 
binaries. Speculatively, these clusters could have 
different orbital period distributions, with more
short period binaries among the Hyades, which would 
show larger coronal activity.
\end{abstract}


\section{Introduction}

Open clusters play a key role in the understanding of
different  time-dependent stellar properties
 such as the evolution of  rotation,
 stellar activity,  and   the lithium abundance.
The comparison between different clusters which have the same age 
allows us to prove if this approach is correct
 or if other effects, such as e.g. a 
different metal content, are also important.
In this report, we examine the Hyades and Praesepe clusters.

During the last 15 years, it has been demonstrated that X-ray
 emission is a `normal' characteristic of late type stars. As 
with other stellar properties which depend on rotation
(in this case through the dynamo effect, Parker 1955), 
the emission strength decays with age, as  shown by the
comparison between open clusters of different ages  such as
the Pleiades (Caillault and Helfand 1985;
Micela et al. 1985; Micela et al. 1990; Stauffer et al. 1994),
%
%
and the Hyades  (Stern et al. 1981; Pye et al. 1994; Stern et al.
1994;  Stern et al. 1995).

The Hyades cluster has been extensively studied at X-ray wavelengths. 
The ROSAT
All-Sky Survey (RASS) detected members 
down to Log~Lx=1--2$\times$10$^{28}$ erg~s$^{-1}$, with detection
rates of 90\% for spectral type dG, 40\% for dK and 30\% for
dM  stars (Stern et al. 1995). 
They also found that X-ray luminosity functions (XLDF) of K and
M-type dwarfs are significantly
affected by the presence of a large number of binary systems in
the cluster,
in the sense that an
important fraction of the stars with strong X-ray emission were
binaries.

A   study of the X-ray properties of Praesepe 
has been carried out by Randich and Schmitt (1995). The Hyades
and Praesepe have  
 similar age and metallicity although 
the Hyades are slightly more metal rich.
Moreover, their kinetic properties are quite close  (Eggen 1992) and they
could have  been  born in the same molecular cloud.  Randich and
Schmitt (1995)  presented the results from ROSAT PSPC Raster
Scan images in a  4$^\circ$$\times$4$^\circ$ region. Their
detection rates for Praesepe
were 33\%, 14\% and 13\% for dG, dK and dM stars, respectively.
As a consequence, the X-ray luminosity functions of Praesepe
in each spectral range are
dominated by the upper limits (UL). 
Since a large fraction of the Praesepe Raster Scan was characterized by
a sensitivity similar to that of the Hyades RASS observation, the difference
in the detection rates means that the bulk of the Praesepe population
is underluminous in X-rays with respect to the Hyades.

The goal of this papers is to try to disentangle this problem,
looking for possible reasons of the disparate
behavior of these coeval clusters in X-rays.

We present the Praesepe 
data studied here and the reduction process in Section 2,
where in Section 3 we analyze the data and perform a comparison
with the  Hyades cluster. Section 4 contains 
the more important conclusions derived  from this study.

\section{Observations and Data Reduction}

Observations of Praesepe members analyzed in this work were
obtained in two  observing runs at the MMT. 
The first one, devoted to the study of Praesepe dF-dK stars,
 was carried out on January 11-13, 1995. We used 
an echelle spectrograph, a 1'' slit and 
an intensified reticon detector. 
With this combination we got 
a free spectral range of 5165--5210 \AA{ } and
a spectral resolution of 0.2 \AA,
 which allows us to obtain accurate
radial velocities ($\sim$1 km~s$^{-1}$).
 The second run took place on 
January  27-28, 1996. In this case, 
we used the spectrograph at the  Cassegrain focus,
 a 1" slit, and a 1200 lines~mm$^{-1}$ grating, 
which yield an spectral  resolution of
 $\sim$1.5 \AA{} and a spectral range of   6110-7025 \AA.
We were able to measure rough radial
velocities ($\sim$5 km~s$^{-1}$), H$\alpha$
 equivalent widths (EW) and color indices (see next section).
Our targets were Praesepe dM stars.
In addition, we checked the radial velocities of several stars during 
another run in December 5-6, 1997. In this case, we used the Keck I
 telescope and the HIRES spectrograph. The spectral range covered is 
6300-8725, with a resolution of 0.13 \AA{ } (as measured in a ThAr 
comparison lamp).

The targets were selected following two different criteria. For
the first
campaign we selected Praesepe stars from the Klein-Wassink
(1927) catalog.
These stars have dF-dK spectral types and we selected 
both a group of stars which were detected in X-rays
and a group of stars with the same colors
 but with only  upper limits in  X-rays.
The stars with X-ray upper limits were primarily weighted
towards stars in the same color range as the X-ray detected 
stars, but for which no accurate published radial velocity existed.
 In total, we observed 37 stars with colors in the range
 0.40$^{\rm m}\le$(B--V)$\le1.11^{\rm m}$, which
corresponds  roughly to F4-K5 spectral types (Bessel 1979).
These stars are listed in the first column of Table 1, using the
Klein-Wassink identification number (1927), whereas columns 2, 3,
4 and 5 give the 
cross identification with the lists provided by 
 Jones and Stauffer (1991),
Jones and Cudworth (1983),  Artjukhina (1966a,b)
 and Hambly et al. (1994). 

The M dwarfs sample was selected from Hambly et al.
(1995), and contains  38 stars with colors in the range 
1.11$^{\rm m}\le$(R$_{\rm R65}$--I$_{\rm N}$)$\le2.11^{\rm m}$,
which corresponds to spectral types dM0--dM5.

In order to perform different calibrations, such as the color
calibration and radial velocities corrections,
 we also observed in all three campaigns a sample of nearby 
late type stars (Gliese 1969).

The X-ray luminosities were retrieved from Randich and Schmitt
(1995) or, in the case of the upper limits, from Randich (private
communication).


\section{Analysis and Discussion}

\subsection{ Membership for dF-dK stars in Praesepe}

One of the goals of this paper was to establish if the available
catalogs of Praesepe stars include a substantial number of
non-members, which could  bias the interpretation of any
statistical study in the cluster. In particular, we intend to
verify if the X-ray characteristics  of this cluster are
affected by the presence of spurious members. 

\subsubsection{ A color-magnitude diagram for dF-dK stars}

     Figure 1 shows the stars studied here (solid symbols) and
other
Praesepe stars (open circles). Stars which have been detected in
X-rays are plotted as solid circles, whereas those
 having only  upper  limits are shown as solid triangles.

     This diagram allows us to establish two different facts:
First,  all of the  stars studied here are, from the photometric point of
view,  possible members of the cluster.
By definition, they are proper motion members since they were 
selected as such from  the Klein-Wassink (1927) catalog.
 Moreover, most of them have been studied recently in
 order to get new measurements of their proper motions
(Jones and  Cudworth 1983; Jones and Stauffer 1991; Hambly et al.
1995)  and these studies  agree in the membership.
(See proper motions and membership probabilities in columns 6--12
of Table 1.) Second, some of these stars can be classified
 as probable  photometric binaries
(PhB), because of their position on the color--magnitude diagram.
Adopting a displacement greater then or equal to  0.4 magnitudes above
the single star Main Sequence as a criterion for duplicity,
 we have found that 14 out of 40 stars are binaries. 
Nine of these 14 binaries 
have been  previously identified as spectroscopic binaries
(Stauffer et al. 1997; Mermilliod 1997b).
Of the remaining five, 
our radial velocity measurements (next subsection) show that 
one is a non-member and that another one is SB.
If we select only the sample detected in X-rays, there are 9 PhB
out of 16 (69\%).
 In the case of the stars with X-ray 
 upper limits, there are 3 PhB
and 20 supposedly 'single'  stars (13\%).

\subsubsection{ The radial velocities for the dF-dK stars}

The radial velocities of our two samples of  dF--dK Praesepe
stars were derived from cross-correlation.
Based on our previous experience with radial velocities derived from
the MMT echelle (Stauffer et al. 1994; 1987), the radial
velocities should  have one sigma accuracy of better than 
1 km~s$^{-1}$. The uncertainties on Keck radial velocities are of the same 
order.
Some of our stars were observed twice 
and three times, in order to 
verify the stability of our measurements and to look for orbital
motions.
Table 2 lists the stars in the first column. The second one
contains the final velocities, whereas the third lists the date
of the
observation in JD. The fourth and the fifth columns
 contain the visual magnitudes and the (B--V) 
color indices, which were selected from Johnson (1952),
Upgren et al. (1979) and Weis (1981).  
In the sixth column we give the X-ray luminosity. 
 Finally, the seventh column contains
 an assessment about the binarity: 
we indicate if the star is a photometric binary and/or a
spectroscopic binary (based on the variations 
of the radial velocities). 
Those stars having several different  measurements of their
 radial velocity  have been  classified as 
 spectroscopic binaries.
The average value of the radial velocities presented here is
37.7 km~s$^{-1}$,
which is in excellent agreement with the average radial 
velocity of all data published to date 
($<$rv$>$=34.5$\pm$1.7 km~s$^{-1}$). Individual radial velocities
also agree quite well with the measurements obtained by Mermilliod
(1997b). Radial velocities show that, essentially, there are not
a significant number of spurious members in our sample of dF-dK stars.

\subsubsection{ Membership and spurious members in dG Praesepe  stars}

In order to make a further search for non-members in
the Praesepe catalogs, we have performed an additional test. We have selected
{\it all} Praesepe presumed members in the color range 0.50$\le$(B--V)$<$0.80
(Klein-Wassink 1927, Artjukhina 1966, Jones and Cudworth 1983, 
Jones  and  Stauffer 1991,  Hambly et al. 1995). We gathered all the available 
information about the BV photometry, proper motion, radial velocity (including 
the unpublished data by Stauffer et al. 1997 and Mermilliod 1997b) and 
lithium abundance (Soderblom et al. 1993). Each star was classified as a member
or non-member based on each of these measurements. The final classification was
 based on all the data. Table 3  lists the
star's names, V magnitude, (B-V) color and  X-ray luminosity in the first 4
 columns.  The last columns contain other identifications.
Statements about the membership to the cluster can be found in 
columns 5 to 9 (the flag ``Y$_{\rm ph}$''
 identifies possible photometric binaries and  ``Y$_{\rm SB}$''
denotes  known spectroscopic binaries). Stars having probability 
membership based on the proper motion larger than 0.9 are singled out with 
a ``Y'' letter in column 6, whereas stars with probability
 between 0.5 and 0.9 appear as ``Y-''. In this last case,
if either the information about the lithium abundance or the 
radial velocity was missing (in very few cases),
 we assigned a ``Y?'' for the final membership (column 9),
 meaning possible member. Stars having ``Y'' are  probable members.

This list of dG stars includes 16 stars which were not included in the
raster scan area and thus do not have X-ray emission measurements.
 In the following discussion, we have not taken
 them into account. We have found that all dG stars having Lx detections
(24 stars) are probable or possible members, using our definitions,
which are quite restrictive.
 In fact, there are only two stars which are just ``possible'' members, due
to the low membership probability, although the radial velocities agree
with the average value of the cluster. In any case, a minimum 91.7\% are
bona fide members of the cluster.  In the case of the stars having Lx upper 
limits (50 stars), we have detected two stars that are  non-members
 (KW258, and JS83), 
since their radial velocities are different than the average of the cluster
(Mermilliod 1997b). There are other 7 stars which are just ``possible''
 members. The rest, 41 in total (82\%), are probable members.

Since as a group,  the complete sample of
 Praesepe dG stars having  X-ray upper
 limits (as well as the stars having detections) 
fulfills the photometric, the proper motion, radial velocity and 
lithium abundance  criteria
 to be members,   we rule out the possibility of a significant
contamination by non-members in the complete sample  of dG stars 
(as we have  done previously  with our sample of dF-dG stars) 
and we suggest that this result can be extended to those 
stars included in the study by Randich and Schmitt (1995).
Further high resolution spectroscopic observations,
 together with  other tests such as lithium abundances and
rotational velocities could confirm the fact that there is not
contamination by spurious members of the sample studied by
Randich and Schmitt (1995).

\subsubsection{ Re--analyzing the Lx data  for dG Praesepe stars}

Figure 3a,b shows the X-ray luminosity against (B--V) color index
for dG Hyades and Praesepe stars, respectively
(0.5$\le$(B-V)$<$0.8). The data are from
Stern et al. (1995) and Randich and Schmitt (1995). Detections
appear as solid circles, whereas upper limits are shown as open
triangles. The horizontal lines represent the location of the
lowest Hyades and Praesepe detections (Log Lx $\sim$ 28.4 and
28.8 erg\,s$^{-1}$, respectively).

Since we have shown that the Praesepe sample is not contaminated
by non--members, the differences in the Lx distribution are, in
principle, real. In particular, the high number (13) of Praesepe members 
having upper limits below Log Lx=28.4 is remarkable ,
when  there is only 1 Hyades member in this region. Randich and
Schmitt (1995) stated that the sensitivity in the inner region of ROSAT PSPC
Raster Scan observations was similar to the  Hyades ROSAT 
All--Sky Survey analyzed by Stern et al. (1995). 

The most obvious explanation for the discrepancy between Lx
properties of  the dF--dK stars in the Hyades and Praesepe clusters
is that the sensitivity of the Praesepe observations is worse than
what was claimed by Randich and Schmitt.
Starting from the hypothesis that the Praesepe XLDF for G dwarfs is
indeed the same as for the Hyades, we estimate that, in order to get
a detection rate of 33 \% in Praesepe, one would need a sensitivity threshold
$\log$ Lx $\sim$ 29.15 erg~s$^{-1}$. 
This, on turn, would mean that Randich and Schmitt (1995) underestimated
ULs by a factor of less than 2 for two of the G dwarfs in their sample,
by a factor 2--5 for nine of them, by a factor 5--10 for eleven of them,
and by a factor $>$ 10 for eight of them.
Whereas ULs could indeed have been underestimated by a factor up
to 2 (see below), it is difficult to think that they could be in error
by larger amounts. 

In order to further check on the ULs issue
we carried out an additional test.
We retrieved from the ROSAT archive a $\sim$8,000 seconds
pointing covering part of the Praesepe cluster. This pointing
includes 12 dG stars undetected in the Raster Scan observations
analyzed by 
Randich and Schmitt (1995). Six of them have cluster membership 
probability  higher than 90\%.  We did not detect any of these
sources in this pointing; this  is not surprising since these
observations were not as deep as the Raster Scan, but it confirms
that there are no major problems with the raster analysis itself.
We determined Lx upper limits
and  found that the {\it new} upper limits are consistent with those
from Randich and Schmitt (1995). These last values were,
on average, a factor  2 lower (0.3 dex), which is not enough to
explain the differences in the detection rates for both clusters.
Therefore, other explanations should be considered.


As a final consideration, we note that the distance to Praesepe as
determined from Hipparcos is larger than what was previously estimated
(e.g., Mermilliod et al. 1997). However, the difference is below
15 \%, and thus not enough to explain the discrepancy in the X-ray
results.

\subsubsection{ Are  rotation and binarity the explanation of the
differences in the X-ray properties of the Hyades and Praesepe?}

One possible explanation of the difference in the X-ray properties
of dK stars the Hyades and Praesepe would be 
 that they have different  rotational velocity distributions,
 which would produce differences in
the coronal and chromospherical activity. Stern et al. (1995) have shown 
that, for stars redder than (B--V)$\sim$0.8, there is a correlation 
between Log Lx/L$_{\rm bol}$ and  rotation. If the difference in
the X-ray luminosity functions of the Hyades and Praesepe is produced by a 
difference in the distribution of rotation, this  fact would
challenge our understanding of the evolution of the angular
momentum, since rotation depends, essentially, on age.
However, Mermilliod (1997a) has
shown that the distribution of rotational velocities is quite the
same in the Hyades and Praesepe F and G-type dwarfs, 
and it is reazonable to assume that this situation holds for the dK stars.
Therefore, a difference in the distribution of rotation is
 very unlikely to be the source of the Lx dichotomy.
An alternative possibility would involve binary systems.

Nine stars of our sample have been identified as spectroscopic 
binaries by Mermilliod (1997b) and Stauffer et al. (1997).
Our observations and those from Mermilliod (1997b) show that
KW460 is, probably, a non-member. Our RVs and the proper motions show that
JS83 is  also a non-member.
Using the radial velocity as criterion, we have classified an  
additional 5 Praesepe members
out of these 38 stars as spectroscopic  binaries (their RVs are
variable and they fulfill
other membership criteria). Among these 14 spectroscopic binaries, 9
were detected in X-ray   and 5 have only upper limits. 
It is worth  stressing that KW478 only has a single
 measurement of the radial velocity. It has been classified as possible SB,
but it could be in fact non-member. However,
since its photometric properties and proper motion
(99\%) agree with it being a member,  we strongly
 believe  that it is indeed a binary. In any case, its membership, 
positive or negative, does not affect our conclusions.
 
Table 4 contains the binary percentages over the total number
of stars (binaries plus singles) of different groups 
in Praesepe and the Hyades. Columns 2 and 3 list the quantities
obtained using the color-magnitude diagram and the radial 
velocities alone, whereas column 4 shows the final
percentages obtained by combining the previous 2 columns. The
last 
column in the table lists the percentages of known
binaries in the Hyades cluster.


Visual inspection of Table 4  shows that the Hyades and Praesepe
have approximately the same percentage of binaries in the three
categories (total number, Lx detection and upper limits), although
the percentage of  total binaries is slightly lower in Praesepe
than in the Hyades.
However, Mermilliod (1997a) has obtained a smaller binarity rate
for Praesepe (34 \%), using a larger sample (85 stars) in the color
range (B--V)$\sim$0.4--1.0. Since binaries tend to be more active 
than single stars, this situation brings one possible explanation
 for the differences in the  luminosity functions
of Praesepe and the Hyades: that they are due to a different binarity
rate  in both clusters, at least when referring to dF-dK stars.
Since Praesepe has less binaries, its Lx emission should be, 
on average, less than in the case of the Hyades. However, the
binarity rate in Praesepe is still not very well known,
contrary to what is true for the Hyades,  and this fact
makes it uncertain that  a difference in binary
rates is the  explanation for the difference in the XLDF.

It is also interesting to compare the total  detection rates
for binaries. We have gathered all the available information 
about binarity (either photometric or spectroscopic) in the
color range 0.40 $\le$ (B--V) $\le$ 1.11 for all known Praesepe
stars. This color interval approximately  corresponds to  
 our sample of dF-dK Praesepe stars. In this color range, there are
59 binaries. Seven have no information about X-ray luminosity, 30
have upper limits and another 22 have been detected by Randich and
Schmitt (1995). Thus,
the detection rate of the binaries in Praesepe cluster is 42\%. 
 However, in the case of the Hyades, Stern et al. (1995) detected
 80\% of the binaries. If only our color range is considered,
they detected  96\%. Therefore, there is an important 
difference in the binary detection rates for both clusters. Even if there 
are unknown binaries in  Praesepe (essentially, all the binaries
in the Hyades have already been identified), an important fraction
of those known binaries are X-ray underluminous when compared with the 
Hyades. It could be argued that some of the Praesepe single stars
with X-ray detections are in fact binaries (therefore, the actual
binary detection rate would be higher). However, Table 2 shows that these
stars have  radial velocity in good agreement with the cluster (therefore, 
it is unlikely that they are spectroscopic binaries), and they are not 
possible photometric binaries. In addition there are a significant number
of Praesepe binaries with X-ray upper limits, contrary to the situation in the
Hyades. In general, since one expects that there is a close physical
relation between rotation and coronal activity, it seems
 likely that differences in average  rotation rates are causing the 
differences in the observed X-ray properties. 
Another explanation could be differences in the
properties of the orbital elements: 
in principal, the geometry of the binaries does not depend on 
the initial evolution of rotation or the initial spin angular momentum.
If the orbital periods distributions
are different (P$_{\rm orb}$ would be larger in Praesepe), Praesepe
stars could show less stellar activity, since short period binaries
rotate faster, on average, than single stars and long period binaries.
Due to the fact that there  are more single 
stars than binaries, and in  any case there is not need 
of too many binaries in  Praesepe rotating  slower than Hyades binaries,
  we would have simultaneously a similar 
distribution  of rotation and some Praesepe binaries being Lx under-luminous.
  Eccentricity could also play a role.  Extensive and complete 
studies of the  Praesepe binaries, in particular of their orbital
periods, could help to establish the reason of the differences
in the X--ray properties.

Metallicity could affect the level of X-ray emission, as ROSAT observations
of NGC 2516 have shown (Jeffries et al. 1997) and 
therefore the difference in metallicity between the Hyades and 
Praesepe could be responsible for the difference
in the X-ray properties.
However,  Section 3.2.5 shows that the chromospheric activity distribution,
 as measured by the  H$\alpha$ emission line,
  is  quite similar in both clusters, 
proving that a small difference in the metallicity is not an important factor.
Therefore,   we think this is an unlikely explanation for teh X-ray differences
 given the small difference in the Praesepe-Hyades metallicity.


\subsection{ H$\alpha$ for the dM stars in Praesepe}

\subsubsection{ A color-magnitude diagram for dM stars}

As discussed in the above sections,
Randich and Schmitt (1995) showed  that the X-ray
luminosity  functions of dF-dK stars belonging to 
the Hyades and Praesepe do not agree. The same
 work found  that this is also true for dM
stars (30\% versus 13\% in the Hyades and Praesepe,
respectively). In order to  try to understand
these differences
we have observed a sample of 38 Praesepe dM stars, {\it all} of
them being undetected in X-rays.
 These stars were selected from Hambly  et al. (1995).

We have measured  H$\alpha$ equivalent widths and  radial
velocities 
for these stars. In addition, we have estimated  (V--I)$_{\rm C}$
color indices using the MAGBAND  routine inside 
IRAF\footnote{ IRAF is distributed  by National 
Optical Astronomy Observatories, which is operated by
the Association of Universities for Research 
in Astronomy, Inc., under contract to the National
Science Foundation, USA}. This routine provides the total
number of counts in several predetermined spectral bands. We
measured these quantities for three bands,
 namely 6170--6270 \AA, 6500-6550 \AA{ } and 
6790--6850 \AA. Since these bands are strongly affected by the
presence of TiO bands in dM stars, which depend
 themselves on the color, we were able to 
calibrate the ratio  between the fluxes in these bands against 
(V--I)$_{\rm C}$ by using several Gliese stars (Gliese 1969),
and, thus, to obtain these indices for the Praesepe
 M dwarfs. Details about the process can be found  in Stauffer and 
Hartmann (1986). An estimation of the error of the colors is
0.05.

Table 5 lists the program stars in the first column, the
photographic
I magnitudes selected from Hambly et al. (1995) 
in the second, and the estimated (V--I)$_{\rm C}$ colors 
in the third. These data allow us to
build a color-magnitude diagram and to verify if the photometry
of these stars agrees with membership. Figure 2 shows such  
diagram.
Based on it, we have classified the stars of our sample of M
dwarfs as members (when they lie on the average MS,
 solid circles on the plot) and possible members (if they are 
slightly separated from it, open circles).
 Following this criterion, HSHJ~16, 23, 165 and 88 are possible
 members. The first three are fainter than our  Praesepe MS 
and  HSHJ~88 could be a binary, since it is slightly brighter.

\subsubsection{ Radial velocities}

Radial velocities were measured comparing the positions
of different lines with those of several Gliese stars
which have accurate values of their radial velocities,  via cross
correlation techniques and the XCSAO routine inside the IRAF
package RVSAO.  The resolution of these spectra, $\sim$1.5 \AA,
allows us to derive radial velocities which are accurate enough 
to serve as one test of cluster membership.
 to  provide enough accuracy in the radial velocity
to assert the membership status based on this criterion.
 For the stars having H$\alpha$ in emission we used a narrow
spectral region around this line, and this method provides an
accuracy of  $\sim$3.5 km~s$^{-1}$. For other stars, we used other
spectral regions and these measurements rely primarely on 
TiO bands. The accuracy in this case is $\sim$7 km~s$^{-1}$. 
Columns 6 and 7 of Table 5 list the barycentric  radial
velocities and
the dates of the observation.

\subsubsection{ The comparison of the coronal activity}

Figure 4a shows the X-ray luminosity measured with ROSAT (Randich
and  Schmitt 1995) versus our estimated
values of the (V--I)$_{\rm C}$ color index for the dM Praesepe
stars studied in this section.  The upper limits range from
Log~Lx$\sim$27.9 to Log~Lx$\sim$29.1 erg\,s$^{-1}$. 
In the figure, solid
triangles represent those stars
having low values of their H$\alpha$ emission (see next section)
and solid circles those having high values. The stars studied by
Williams et al. (1994) are also  included in the diagram
as open circles. From now on, we will include these data 
in our analysis, since there is no reason to think
 that there is any systematic difference between both
 sets of data.

Figure 4b contains the same kind of data, but the Hyades are shown.
Here, solid circles represent actual values of the X-ray
luminosities whereas triangles represent upper limits. The X-ray
luminosities were selected from Stern et al. (1995).

As pointed out by Randich and Schmitt (1995), the
detection rate is less in Praesepe than in the Hyades, resulting in
different luminosity functions  (see their Figures 8a,b).
There are several interpretations to this fact: 
they considered  the possibility of contamination
 by spurious members,  a real age
 difference between both clusters,  an effect of their
different
metal contents, and differences in the distribution of rotational
velocities.
However, the isochrone fittings give quite similar ages for both
cluster  and Soderblom et al. (1993) have shown that the
 lithium abundance distribution follows the same pattern
 in Praesepe and  the Hyades. Since lithium is extremely 
dependent on age, as suggested by Iben (1965; 1967a,b), as is 
rotation (e.g., Chaboyer et al. 1995),
and we have proven that the sample of 
Randich and  Schmitt (1995) is not likely to be
 contaminated by non-members, it could be that the difference in
metallicity is playing  an important role in the dynamo effect 
or we  should look for another cause.

\subsubsection{ The chromospheric emission}

Figure 5a shows the H$\alpha$ equivalent widths of a sample of
Hyades M dwarfs versus their (V--I)$_{\rm C}$ color indices
--EW(H$\alpha$) from this report and Williams et al. (1994).
Stars having X-ray luminosities are represented 
with solid circles, whereas solid triangles represent stars
 with Lx upper limits. Known Hyades stars with
 no available Lx data are shown as crosses. We have traced
 a solid line on the plot which separate the 
active coronae stars from those which have  upper limits.
 We will use this line in order to
 discriminate two groups of dM stars in Praesepe.

The figure shows clearly that the coronal emission is tightly
related to the chromospheric emission, as measured with ROSAT 
X-ray data and the H$\alpha$ line: for a given color, Hyades stars
have higher H$\alpha$ emission if they have been detected in
X-rays.
This is understandable, since the mechanism that heat the
chromosphere, 
should also be
responsible for the coronal heating. 

Figure 5b is the same as Figure 5a, but in this case we show the Praesepe
sample. 
 We have superposed the line defined in
Figure 5a in this plot, in order to discriminate between  two
groups of Praesepe stars:
the first group is  composed of stars with high values of
their H$\alpha$ emission (and, presumably, relatively high
coronal activities and therefore X-ray luminosities), and the
second one is composed of stars with H$\alpha$ in
absorption or with low emission (and possibly lower X-ray
luminosities than the previous group). These two groups will be
used to  study the statistical properties of the X-ray
luminosities of our sample of dM Praesepe stars against similar
 members of the Hyades cluster.

The direct comparison  between the relation 
EW(H$\alpha$)--(V--I)$_{\rm C}$ for the Hyades and Praesepe (Figure
5a,b) reveals,
contrary to what is expected on the basis of the results by
Randich and Schmitt (1995), that if
there is any difference between this activity indicator  for the
chromospheres in both clusters, this difference is that Praesepe 
dM stars are {\it more active} and not less than the equivalent
Hyades stars. The distribution for (V--I)$_{\rm C}\le$2.3 is
quite the same, but for redder colors a few stars with very
large EW(H$\alpha$)s are present in Praesepe but not in the Hyades.
On the other hand, it is possible to note that
although the Hyades M dwarfs having upper limits are located in a
well defined region in the diagram, the Praesepe stars have a
larger spread in their EW(H$\alpha$) values. In particular, some
of them are quite strong, and larger than those present in
Hyades stars which were detected in X-rays for the same color.

The H$\alpha$ distribution is approximately equal in both
clusters, and  it is a well--known fact that this indicator of
the chromospheric activity  depends ultimately on rotation.
Therefore, the obvious conclusion
is that, as it is the case for dF-dK stars,
rotation is not responsible for the
differences in the Lx properties of  the Hyades and Praesepe M
dwarfs.

\subsubsection{ A statistical comparison with the Hyades}

In order to clarify the differences and similarities in the
activity
properties of the Hyades and Praesepe clusters, we have built four
histogram diagrams. Figure 6a shows the frequency 
(bind width=1 \AA, but the results are not sensitive
to the binning) of the H$\alpha$ equivalent widths for those
Hyades stars having known values of their EW(H$\alpha$)s,
 whereas Figure
6b contains the same histograms for the Praesepe stars studied
here plus the Williams' et al. (1994) data.
The first visual inspection shows that the general shape of the
distribution is
similar. However, Hyades stars show peaks around -0.5
\AA{ }
and 4 \AA, whereas Praesepe stars are clustered around 1.5 \AA{
} and 3 \AA. The differences between the low emission or absorption
group can be
explained in term of the different qualities of the EW(H$\alpha$) 
of both groups of stars:
The Hyades data have higher signal--to--noise ratios and is some cases
higher spectral resolution,  and,
therefore, provide better measurements of weak  features. The
difference for the high emission group is essentially due to the
 differences between the color distribution of both clusters 
--after all, this comparison does  not use  the same number of
stars at the same color for both clusters, and there is a trend
between the
color and EW(H$\alpha$).   The most relevant difference is the
tail in the Praesepe distribution extending to very high values of the
emission in
H$\alpha$. To elucidate the cause of this difference, we have
plotted the
same histograms using stars  
in a narrower color range, namely 2.4$\le$(V--I)$_{\rm C}$$\le$2.8. 
In this particular case, we have included all Hyades members
with measures
for their EW(H$\alpha$), regardless the known or unknown value 
of their X-ray luminosities. Figures 6c,d show that 
the Hyades and Praesepe distributions are
 quite the same. However, an important difference appears
when comparing the stars with higher EW(H$\alpha$) emission.
 As it was shown in the previous section, Praesepe
stars are likely to have higher chromospheric activity, at least
when comparing the dMe stars.

We have repeated the same analysis with the stellar groups 
defined, for Praesepe and the Hyades, in Section 3.2.4 using the 
EW(H$\alpha$) vs. (V--I)$_{\rm C}$ plane (stars having strong and
weak H$\alpha$ equivalent widths). The conclusions are 
the same in all cases.

A direct comparison between the X-ray luminosities/upper limits
and the  H$\alpha$ equivalent widths can give new clues. Figures
7a,b  show the 
relation between both quantities for the Hyades and Praesepe,
respectively.
Hyades stars which have been detected in X-rays
are represented as solid circles, whereas the upper limits are
plotted as solid triangles in Figure 7a. On the other hand, those
Praesepe stars classified as active in the
EW(H$\alpha$)-(V--I)$_{\rm C}$
plane appear as solid circles in Figure 7b, whereas the
 less active are shows as triangles.
Figures 7a,b evidence a large spread in the X-ray luminosities
for a given H$\alpha$ value in both clusters. Note that in the case 
of Praesepe, all points represent upper limits.
 The only clear conclusion is that the distributions
of points for both clusters seem different.


As in the case of the Praesepe dG stars,
we have re--analyzed the Lx upper limits and found that the old
data are consistent with the new values. Some of these points
may have a higher value of the Lx upper limit, since our re-analysis
shows that some of them could have been underestimated up to
a factor of two. However, the general feature of strong H$\alpha$/low X-ray
emission would remain and it should be explained.
The only remaining reason for the observed differences in
the activity indicators would be differences in the binarity
rates (see the  Section 3.1.5) 
and, specifically, on the distribution of their orbital periods.
 An intensive observing program to measure orbital rotational
periods and to find binaries in the Praesepe cluster
 would help to elucidate if this  speculation is correct.

In any case, Figure 7b shows clearly that there are a handful of
Praesepe stars which have, simultaneously, strong H$\alpha$ and
low values of the Lx upper limit. Since our new re--analysis
shows  that some of them could have been underestimated up to a
factor 2 the Lx upper limits, it would be possible that this
increase could explain, at least in part, this phenomenon 
(strong H$\alpha$ together with low Lx upper limit). But even if
this is the case, it should be explained why the magnetic heating
has been significant enough to produce strong  emission from the
chromosphere, whereas the corona of these stars appears to be
inactive.


\section{Conclusions}

We have tried to establish the reasons of the different
X-ray properties of late type stars in the Hyades and Praesepe. 
We have studied two different samples of stars: dF-dK Praesepe
stars having detected and upper limits for their X-ray 
luminosities and dM Praesepe stars which have been not 
detected by ROSAT.

The measured radial velocities for both samples show that 
contamination by spurious members cannot account for the 
differences in the level of coronal activity, since all
stars (but one) studied here, and presumably most of the
 stars in the Randich and Schmitt (1995) sample, are
real members.

Using simultaneously color-magnitude diagrams and the measured
radial velocities, we have discovered new binaries in Praesepe
for the dF-dK stars. 
The comparison of the fraction of binaries in Praesepe and 
the Hyades shows that it could be slightly different in both clusters.
 Since the observed  levels of coronal activity, assumed equal
sensitivity in the observations, are lower in Praesepe, one would
 expect a smaller binarity rate in Praesepe than in the Hyades.
Moreover, we have shown that the detection rate for the
binaries is much higher in the Hyades than in Praesepe. This could be
interpreted as an effect of a difference in the distribution of
the orbital periods in both clusters.

Finally, the study of the statistical properties of the 
H$\alpha$ spectral line for the dM stars in both clusters shows
that in fact Praesepe presents higher chromospheric activity for
this kind of stars than the Hyades. This result is also surprising, 
since none of the  Praesepe M dwarfs were detected in X-ray, whereas 
many of the Hyades M dwarfs are coronally active.
 For this reason, one possible explanation
for the differences in the X-ray properties between both 
coeval clusters, the existence of different distributions of the
rotational velocities, seems unlikely. We have found several
Praesepe dM stars which have a remarkable strong H$\alpha$
emission and very low Lx upper limits, an unexpected situation.

All these data could indicate   the possibility 
of a difference between the sensitivity of the ROSAT
All-Sky Survey for the Hyades and the ROSAT PSPC observations of
  Praesepe.   However, our re--analysis of several
ROSAT pointings shows that the previous assignation of upper
limits is essentially correct (although there is a suggestion 
that the initial estimates for the upper limits were too low
by a factor of two). We propose  differences in the orbital
period distribution as a partial explanation of the dichotomy of the Lx
properties. Extensive studies of different properties which
characterize late-type  stars, such as rotational velocities and
periods, lithium abundances and additional activity indicators
 should be made  in a large variety of open clusters
in order to have a comprehensive perspective of the evolution
of this type of stars. In particular,
 a similar comparison to that performed here with other clusters
of the same age, such as Coma, could also contribute towards an
understanding of the differences in the X-ray properties of
coeval  clusters. New X--ray data from AXAF or XMM could help to
solve this problem.


\acknowledgements

This research has made use of the Simbad Data base,
operated at CDS, Strasbourg.  J-C. Mermilliod kindly provided
unpublished radial velocities. DBN thanks the ``Real Colegio
Complutense at Harvard University'' and the Commission for Cultural,
Educational and Scientific exchange between USA and Spain for
their fellowships. JRS acknowledges support from NASA Grant
NAGW-2698.


\normalsize

\clearpage  

\begin{center}
{\sc Figure Captions}
\end{center}

\figcaption{ Color-magnitude diagram for Praesepe dF-dK stars: 
apparent visual magnitude against (B--V) color. Solid symbols
represent  the stars analyzed here (circles for stars which  have
been detected in X-ray, triangles for the
upper limits), whereas the the open circles show the position of Praesepe
stars not included in the X-ray survey. 
\label{fig1}}

\figcaption{ Color-magnitude diagram for the Praesepe dM stars: 
apparent visual magnitude against the cousin (V--I) color. Solid
circles represent real members and open circles possible members
based on this diagram.
\label{fig2}}

\figcaption{ Logarithm of the X-ray luminosities against the 
(B--V) color for dG stars. Solid circles and open triangles 
represent Lx detections and upper limits, respectively. The
horizontal lines indicate the minimum value of a detection for
the Hyades  and Praesepe, Log Lx $\sim$ 28.4 and 28.8 erg\,s$^{-1}$,
respectively.  {\bf a} Hyades data.
{\bf b} Praesepe data.  
\label{fig3}}

\figcaption{ Logarithm of the X-ray luminosities against the 
(V--I)$_{\rm C}$ color for dM stars. {\bf a} Praesepe data. Solid
and open
symbols represent the sample observed here  and the stars studied
by Williams et al. (1994). Circles and triangles  correspond to
stars with high and low values of their EW(H$\alpha$),
respectively.
{\bf b} Hyades data. Solid circles  represent the positions
of stars having  actual values of their X-ray luminosity. Solid
triangles show the data having only upper limits.  
\label{fig4}}

\figcaption{ H$\alpha$ equivalent widths against
 (V--I)$_{\rm C}$ color indices for M dwarfs. 
{\bf a} Hyades data. Solid circles and triangles represent stars
having actual values of the  X-ray emission and upper limits,
respectively.  The solid line separates the two groups.
 Stars with unknown X-ray data are shown as crosses.
{\bf b} Praesepe data. The line defined previously, using Hyades
data, is superposed. Circles represent those stars over this
line and triangles those under it. The solid symbols denote 
those stars studied here whereas the open symbols correspond to
the stars studied by Williams et al. (1994). Note that all
Praesepe stars have upper limits for their X-ray luminosities.
\label{fig5}}

\figcaption{Histograms for the  equivalent widths of H$\alpha$
for the dM stars of the Hyades ({\bf a,c}) and Praesepe
 ({\bf b,d}). {\bf c} and {\bf d} show stars in the range 
2.4$\le$(V--I)$_{\rm C}$$\le$2.8.
\label{fig6}}

\figcaption{Logarithm of X-ray luminosities against the H$\alpha$
equivalent widths. {\bf a} Hyades data. Symbols are as in Figure
4b. {\bf b} Praesepe data. Symbols are as in Figure 4a.
\label{fig7}}

\newpage

\begin{figure*}

\vspace{18cm}

\includegraphics{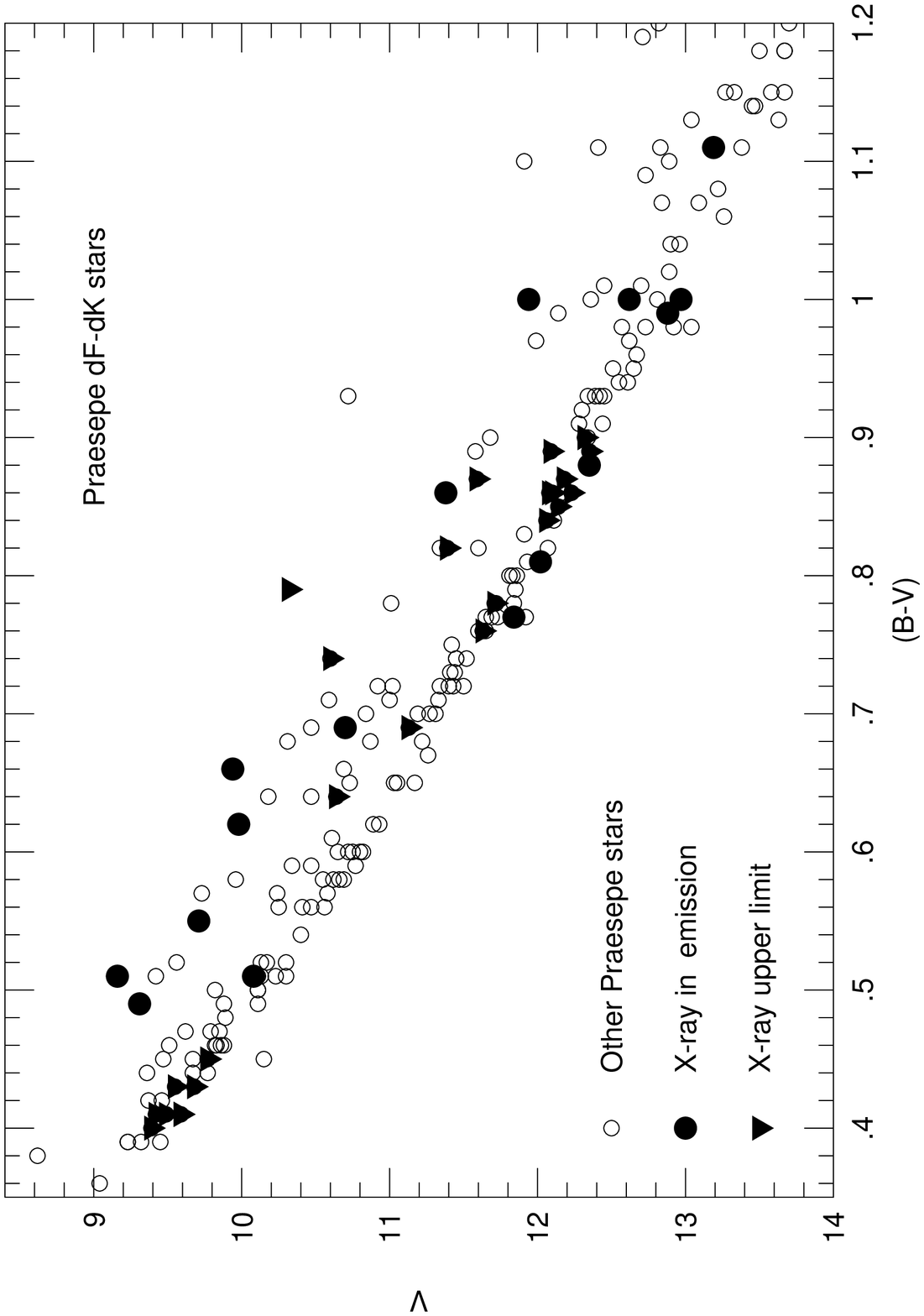}

\end{figure*}

\begin{figure*}

\vspace{18cm}

\includegraphics{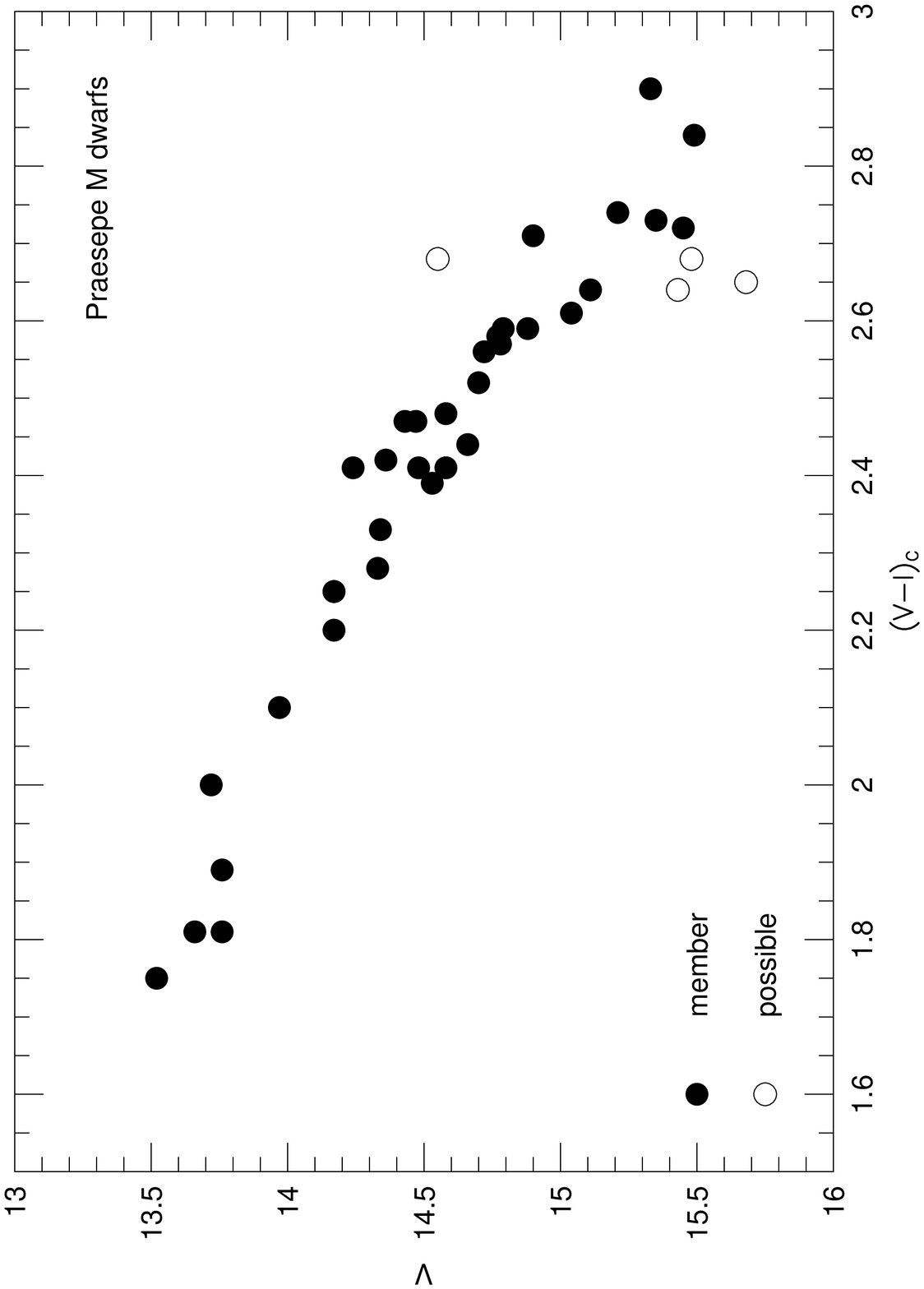}

\end{figure*}

\begin{figure*}
\vspace{18cm}

\includegraphics{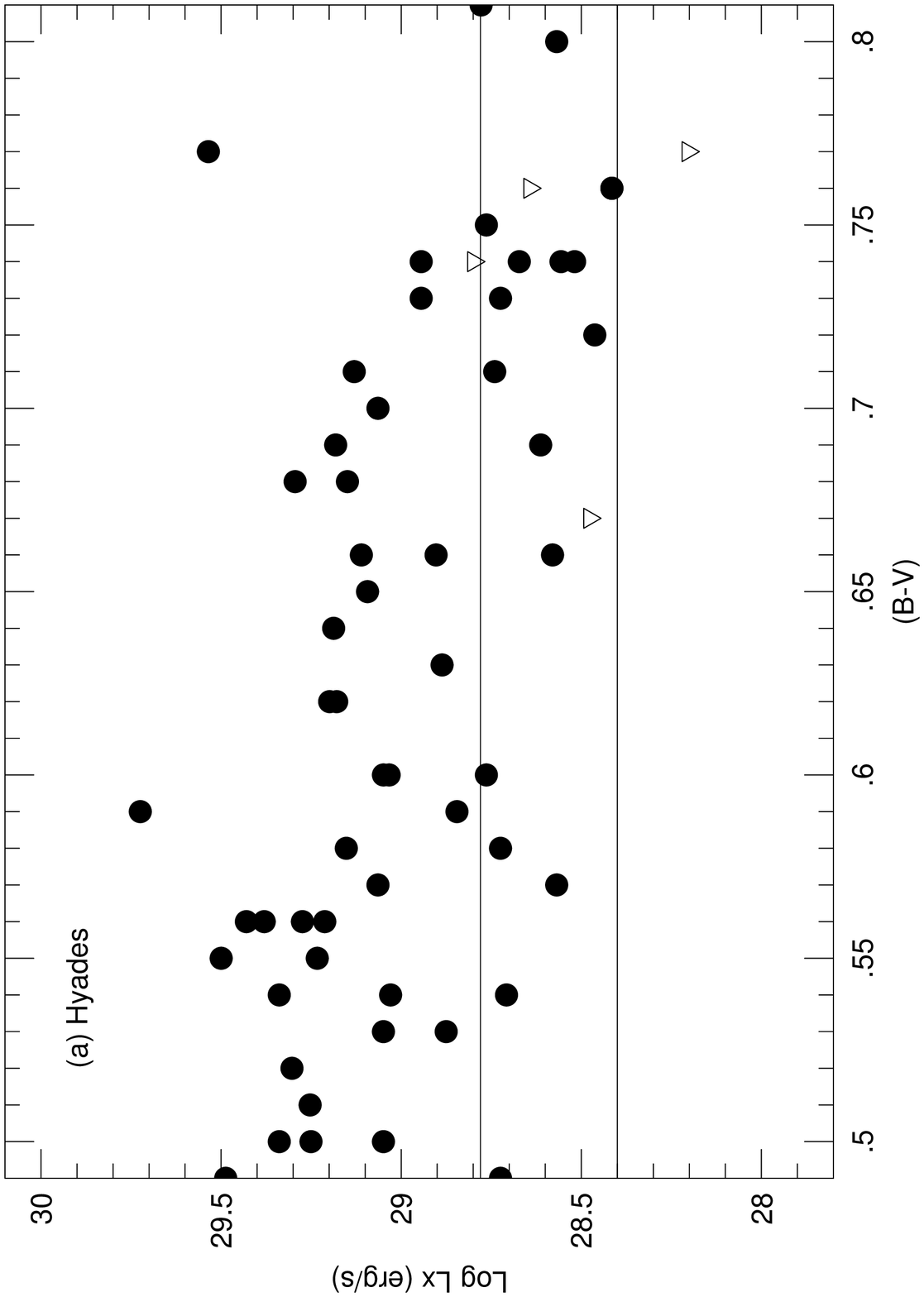}

\end{figure*}

\begin{figure*}
\vspace{18cm}

\includegraphics{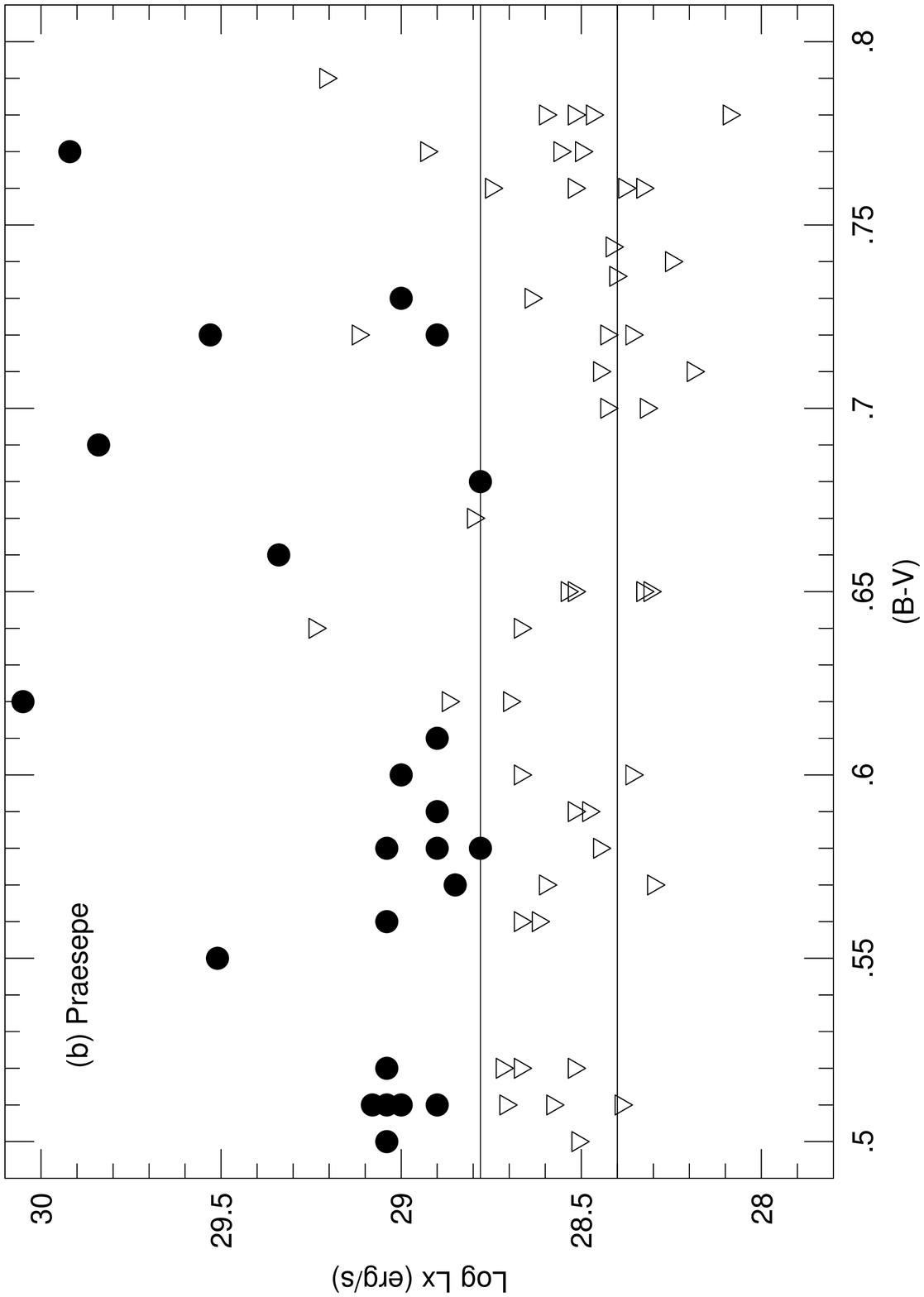}

\end{figure*}

\begin{figure*}
\vspace{18cm}

\includegraphics{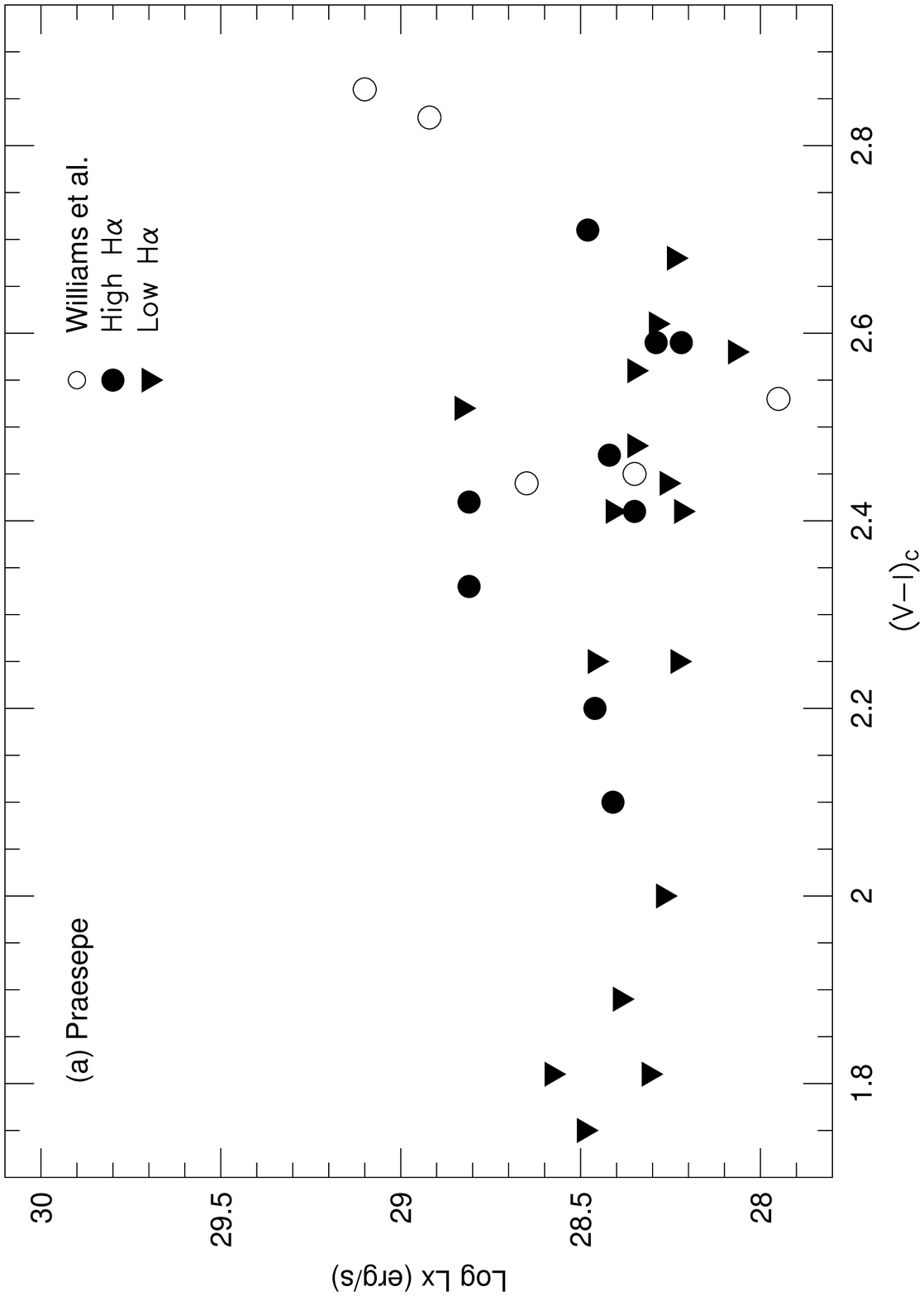}

\end{figure*}
\begin{figure*}
\vspace{18cm}

\includegraphics{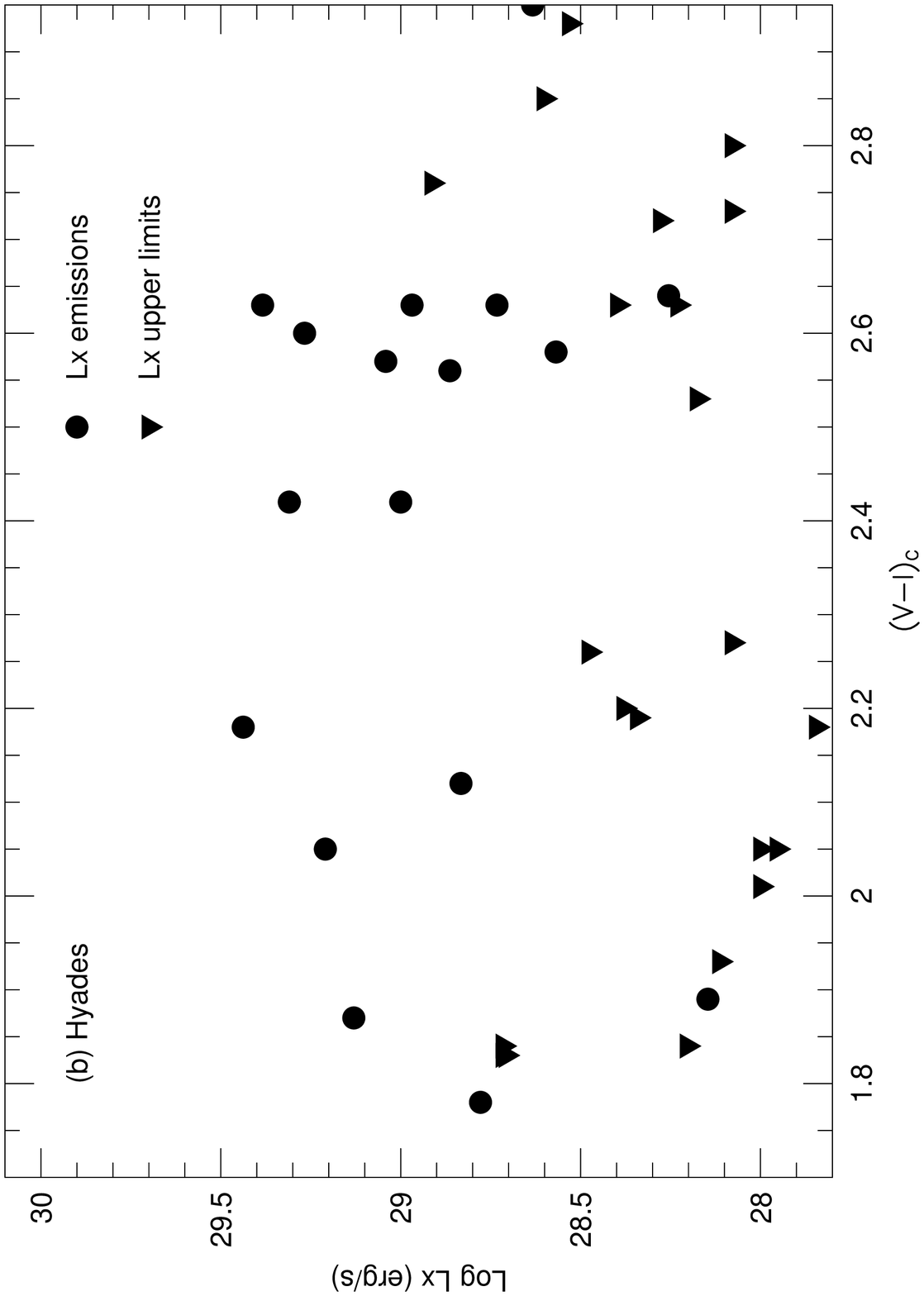}

\end{figure*}

\begin{figure*}
\vspace{18cm}

\includegraphics{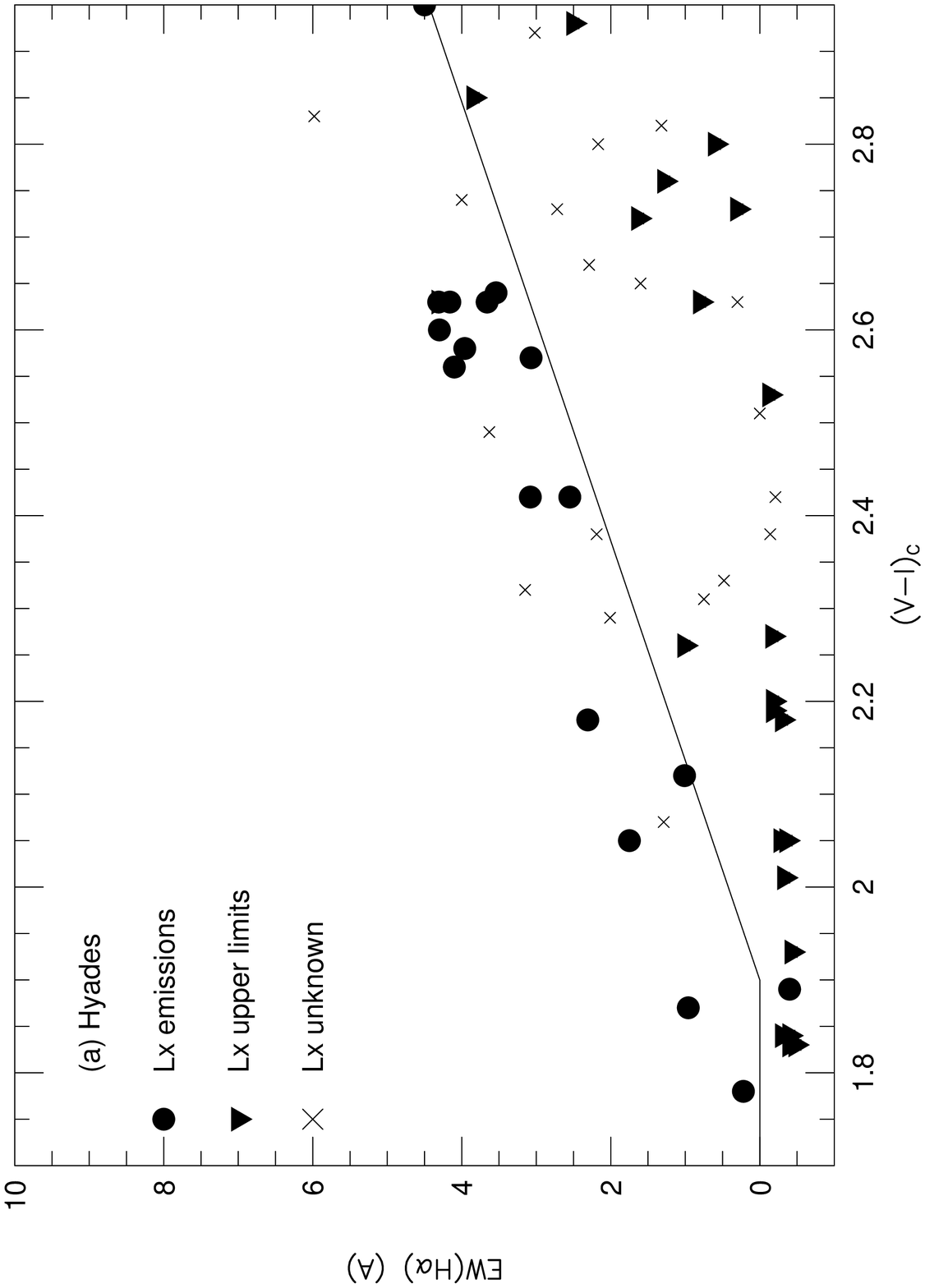}

\end{figure*}

\begin{figure*}
\vspace{18cm}

\includegraphics{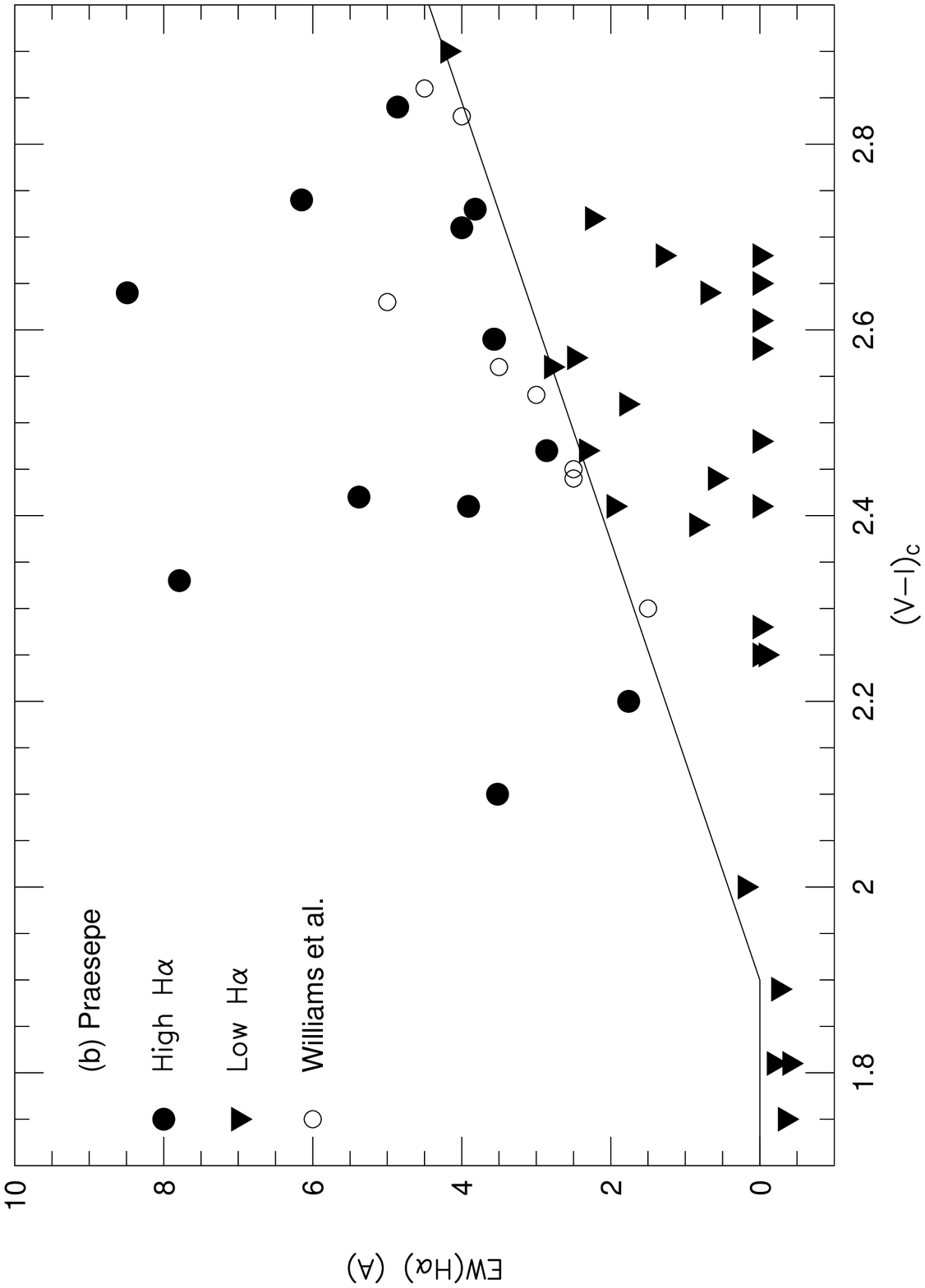}

\end{figure*}

\begin{figure*}
\vspace{18cm}

\includegraphics{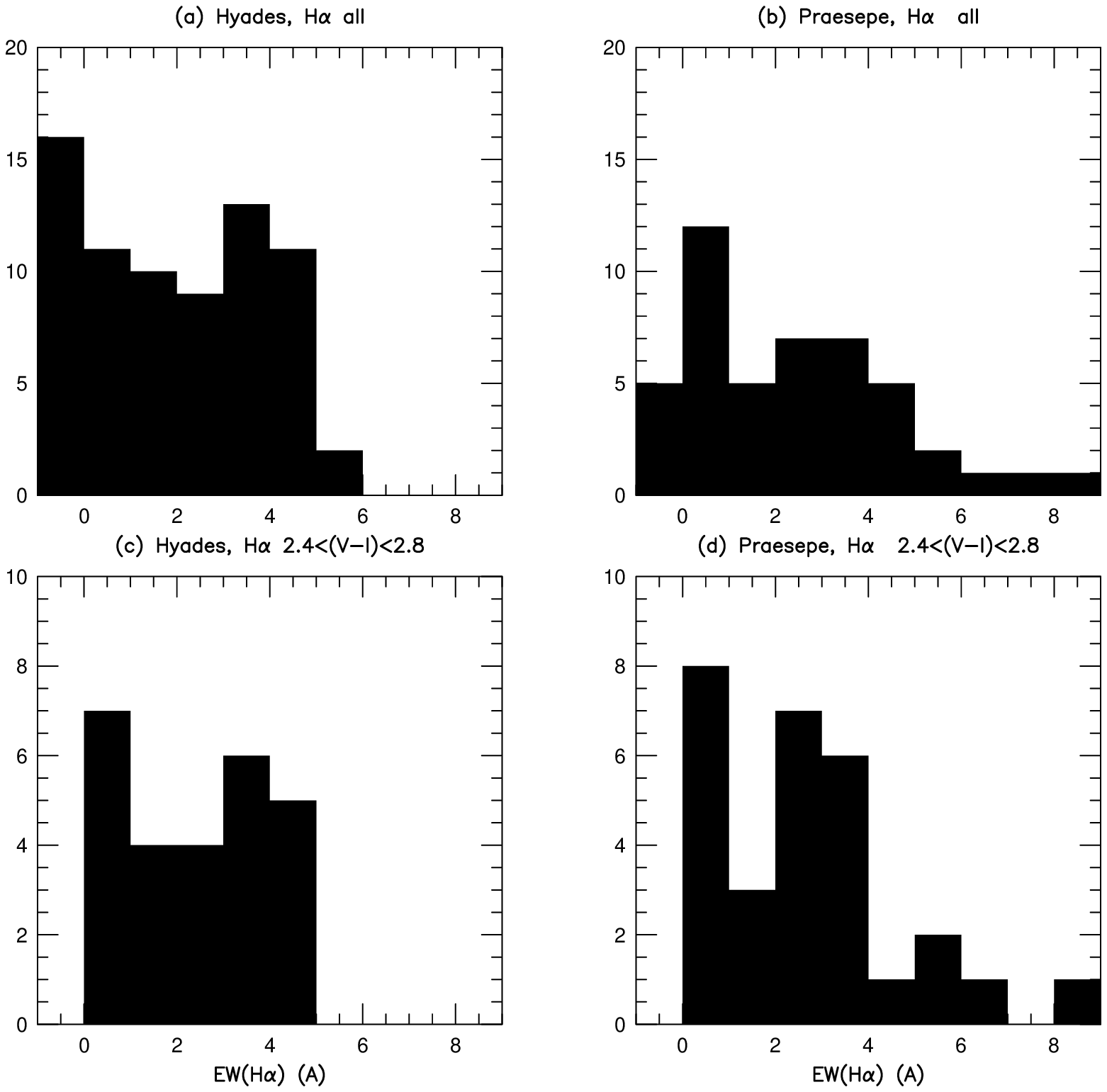}

\end{figure*}

\begin{figure*}
\vspace{18cm}

\includegraphics{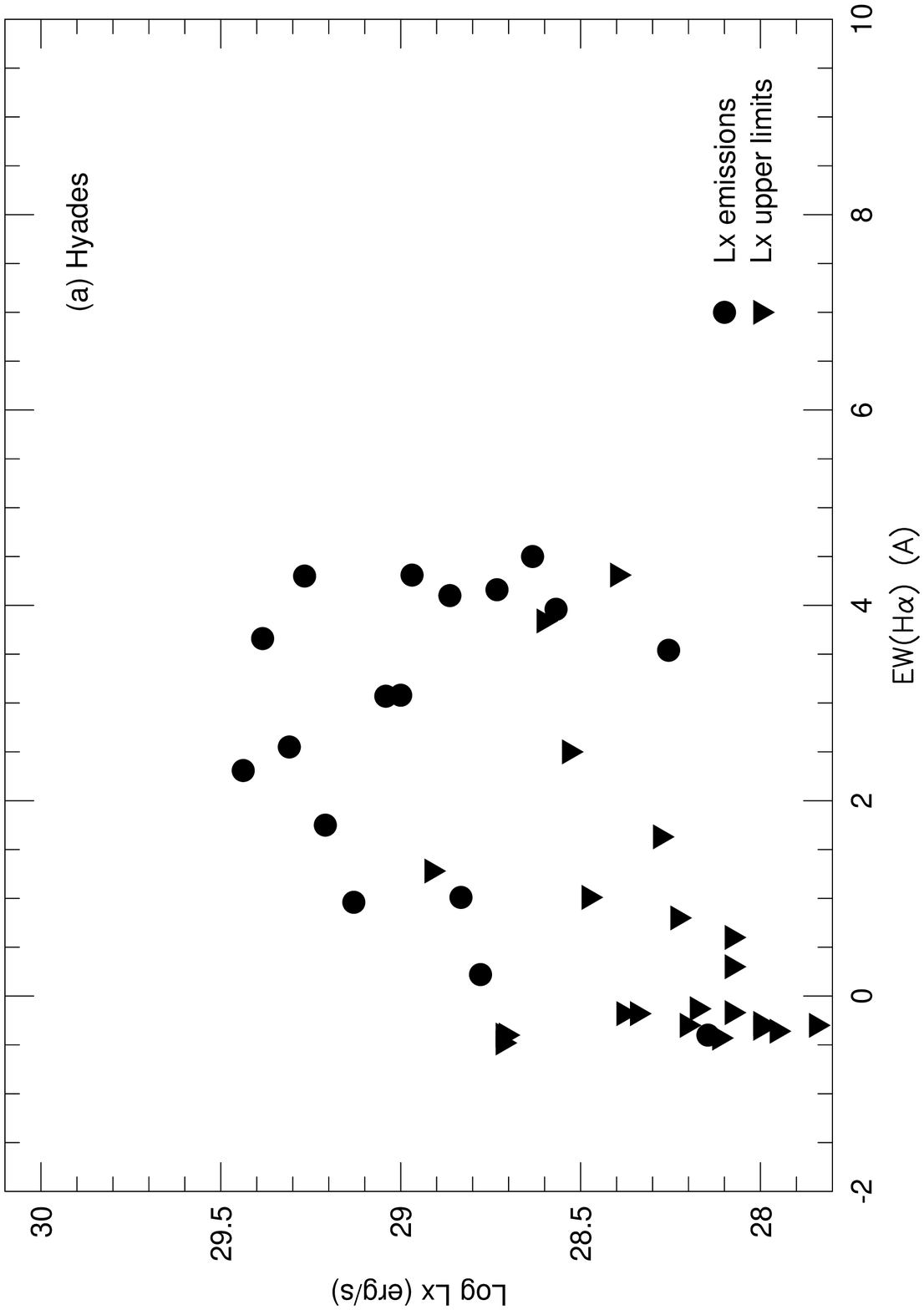}

\end{figure*}

\begin{figure*}
\vspace{18cm}

\includegraphics{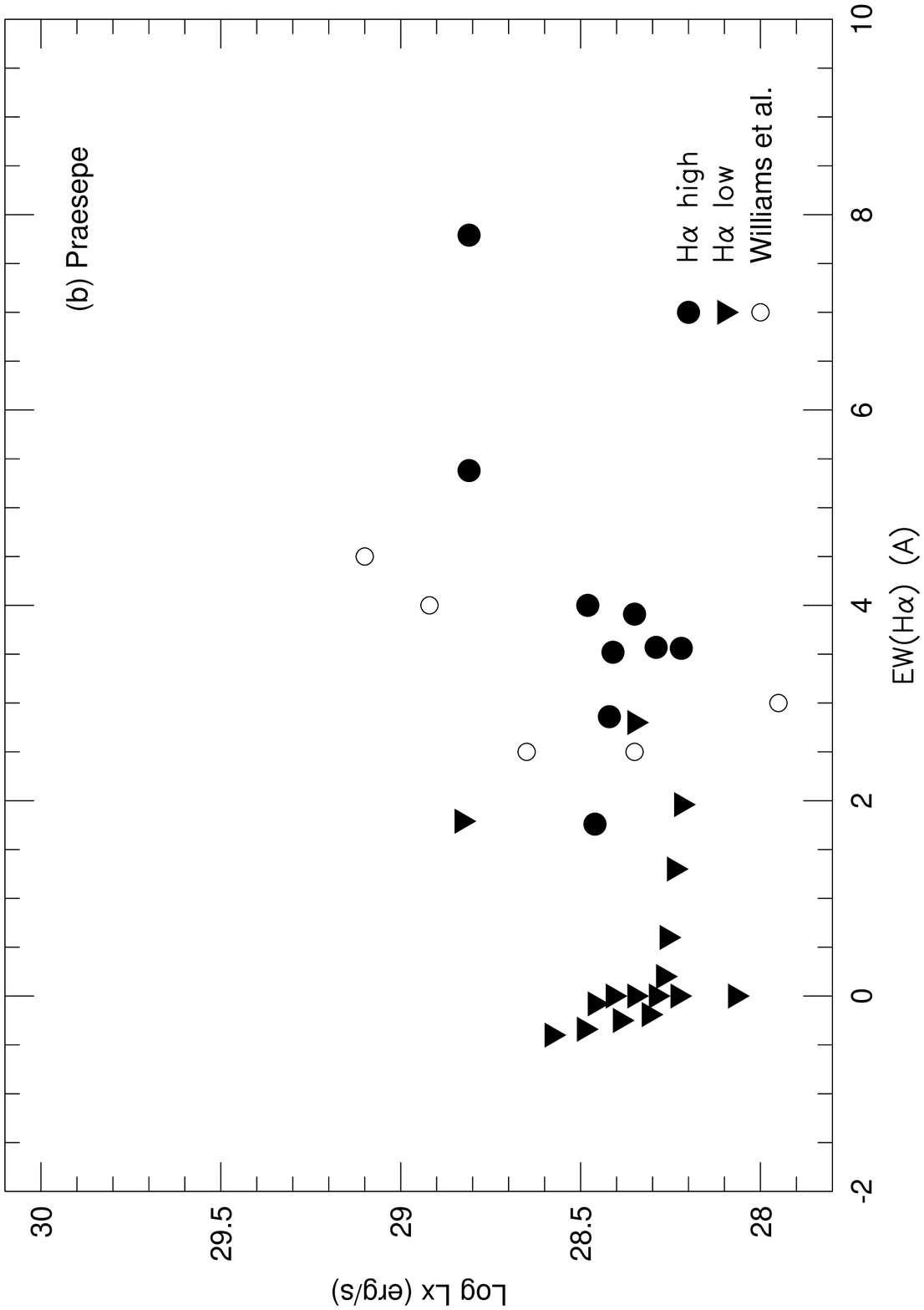}

\end{figure*}

\end{document}